\pdfoutput=1
\documentclass[11pt]{article}

\usepackage[preprint]{acl}

\usepackage{times}
\usepackage{latexsym}
\usepackage{amsmath, amssymb}
\usepackage{subcaption}
\usepackage{graphicx}

\usepackage[T1]{fontenc}
\usepackage[utf8]{inputenc}

\usepackage{microtype}

\usepackage{inconsolata}

\usepackage{graphicx}

\title{Dynamic Bayesian Item Response Model with Decomposition (D-BIRD): Modeling Cohort and Individual Learning Over Time}

\author{
  Hansol Lee\textsuperscript{1} \\
  \texttt{hansol@stanford.edu}
  \And
  Jason B. Cho\textsuperscript{2} \\
  \texttt{bc454@cornell.edu}
  \And
  David S. Matteson\textsuperscript{2} \\
  \texttt{dm484@cornell.edu}
  \And
  Benjamin W. Domingue\textsuperscript{1} \\
  \texttt{bdomingu@stanford.edu}
  \AND
  \textsuperscript{1}Stanford University \\
  \textsuperscript{2}Cornell University
}

\begin{document}
\maketitle
\begin{abstract}
We present D-BIRD, a Bayesian dynamic item response model for estimating student ability from sparse, longitudinal assessments. By decomposing ability into a cohort trend and individual trajectory, D-BIRD supports interpretable modeling of learning over time. We evaluate parameter recovery in simulation and demonstrate the model using real-world personalized learning data.
\end{abstract}

\section{Introduction}
\label{sec:intro}

As personalized learning platforms become more widespread, students increasingly encounter assessments that are short, embedded, and distributed over time. These settings produce sparse but longitudinal data, creating new opportunities—and challenges—for educational measurement. The emerging goal is no longer just to estimate ability at isolated time points, but to track how ability evolves over time, both individually and relative to peers.

Item response theory (IRT) provides a principled framework for estimating latent traits such as ability, but traditional IRT assumes ability is fixed within and across assessments. Dynamic extensions relax this assumption by modeling ability as a time-varying stochastic process \citep[e.g.,][]{martin2002dynamic, wang_bayesian_2013, kim2023variational, 8970762, imai2016fast, sun2025bayesian}. However, most existing models treat students independently or borrow strength only through global priors, limiting their ability to capture cohort-level trends.

We introduce \textbf{D-BIRD} (Dynamic Bayesian Item Response model with Decomposition), a fully Bayesian dynamic IRT model that decomposes each student’s ability into two components: a cohort-level trend capturing shared change over time, and a student-specific deviation capturing personalized growth. This structure enables the model to borrow information across students while preserving heterogeneity in learning patterns. We perform posterior inference via P\'{o}lya-Gamma augmentation~\citep{pg}, which enables efficient sampling and calibrated uncertainty quantification for logistic models.

D-BIRD addresses a growing measurement need in personalized education: estimating learning trajectories in a statistically coherent, interpretable way—even under sparsity. By explicitly modeling both shared and individual dynamics, it provides a foundation for learner feedback, program evaluation, and cohort monitoring.

We validate D-BIRD through simulation and empirical analysis. First, we assess parameter recovery and test its key components via ablation. Then, we apply the model to K–12 reading data from a digital learning platform, demonstrating its ability to recover cohort trends and individual trajectories under real-world constraints.

\section{Model Specification}\label{sec:model}

We present D-BIRD, a dynamic IRT model that decomposes latent ability into a shared cohort trend and student-specific deviations evolving over time. Let \( Y_{i,t,j} \) denote the binary response (correct/incorrect) of student \( i \in \{1, \ldots, N\} \) at time \( t \in \{1, \ldots, T\} \) on item \( j \in \{1, \ldots, J\} \). The goal is to estimate each student’s latent proficiency \( \theta_{i,t} \) at each time point. The model is defined as:
\begin{subequations}\label{eq:fullmodel}
\begin{align}
    Y_{i,t,j} &\sim \text{Bernoulli}(\pi_{i,t,j}), \label{eq:ob1} \\
    \pi_{i,t,j} &= \text{logit}^{-1}(\theta_{i,t} - d_j), \label{eq:ob2} \\
    \theta_{i,t} &= \mu_t + \beta_{i,t}, \label{eq:la} \\
    \Delta \mu_t &\sim \mathcal{N}(0, \sigma^2_{\Delta\mu}), \label{eq:mu} \\
    \Delta \beta_{i,t} &\sim \mathcal{N}(0, \sigma^2_{\Delta\beta_i}), \label{eq:beta}
\end{align}
\end{subequations}
where \( \Delta \mu_t := \mu_t - \mu_{t-1} \), \( \Delta \beta_{i,t} := \beta_{i,t} - \beta_{i,t-1} \), and \( d_j \) is the difficulty of item \( j \).\footnote{We assume item difficulties \( d_j \) are known \emph{a priori}, reflecting common practice in operational assessments where items are pre-calibrated and drawn from a stable pool. While D-BIRD can be extended to estimate item parameters jointly, we focus here on ability estimation under known difficulties.}

Equations~\eqref{eq:ob1}–\eqref{eq:ob2} define a Rasch model \citep{rasch}, where the probability of a correct response depends on the difference between ability and item difficulty. Like other dynamic extensions of IRT, D-BIRD embeds this structure within a temporal state-space framework by modeling ability as a time-indexed latent process. In doing so, it fits within a broader class of dynamic linear models \citep{west1985dynamic, west2006bayesian}, where the key modeling choice lies in the prior placed on the latent trajectory.

In discrete-time settings, common priors over ability include AR(1) processes, as in \citet{wang_bayesian_2013, sun2025bayesian}, and Gaussian random walks, as in \citet{martin2002dynamic} and \citet{kim2023variational}, where each student's ability is modeled as a single latent process with a shared innovation variance. Other work such as \citet{8970762} has explored continuous-time priors such as Gaussian processes, which are particularly relevant when modeling irregularly spaced assessments. While these approaches support temporal smoothing, they typically assume a uniform degree of smoothness across individuals and do not separate shared trends from individual deviations—limiting interpretability when comparing student growth to broader cohort patterns.

D-BIRD also adopts a random walk over ability but structures it differently from prior models. Its key innovation of D-BIRD is an additive decomposition of ability into two components (Equation~\ref{eq:la}): (1) a cohort trend \( \mu_t \), shared across all students and capturing group-level change, and (2) a student-specific deviation \( \beta_{i,t} \), representing individual progress relative to that trend. Both components evolve over discrete time as Gaussian random walks with distinct innovation variances: \( \mu_t \) with a shared variance \( \sigma^2_{\Delta \mu} \) (Equation~\ref{eq:mu}), and \( \beta_{i,t} \) with student-specific variances \( \sigma^2_{\Delta \beta_i} \) (Equation~\ref{eq:beta}). This structure allows for heterogeneous smoothness across individuals while situating trajectories within a common temporal reference.

This decomposition allows D-BIRD to be both flexible and interpretable. It accommodates heterogeneity in student-level learning while supporting cohort-based comparisons and population-level monitoring. In doing so, D-BIRD offers a principled framework for measuring learning progress over time—balancing individualized adaptation with shared structure across the student population.

\section{Inference}
\label{sec:inference}

We perform fully Bayesian inference for the model specified in Equation~\eqref{eq:fullmodel}. Let the observed responses be denoted by \( \boldsymbol{y} := \{y_{i,t,j}\}_{i=1,\ldots,N;\,t=1,\ldots,T;\,j=1,\ldots,J} \). The primary latent variables include the cohort-level trajectory \( \boldsymbol{\mu} := \{\mu_t\}_{t=1}^T \) and the student-specific deviations \( \boldsymbol{\beta} := \{\beta_{i,t}\}_{i=1,\ldots,N;\,t=1,\ldots,T} \).

\paragraph{Prior specification.} Initial values follow Gaussian priors: \( \mu_1 \sim \mathcal{N}(0, \sigma^2_{\mu}) \) and \( \beta_{i,1} \sim \mathcal{N}(0, \sigma^2_{\beta_i}) \). Subsequent values evolve via Gaussian random walks:
\[
\mu_t \sim \mathcal{N}(\mu_{t-1}, \sigma^2_{\Delta \mu}), \quad
\beta_{i,t} \sim \mathcal{N}(\beta_{i,t-1}, \sigma^2_{\Delta \beta_i}).
\]
Variance components include:
\begin{itemize}
    \item \( \sigma^2_{\mu} \): initial variance of the cohort trend,
    \item \( \boldsymbol{\sigma}^2_{\beta} := \{\sigma^2_{\beta_i}\} \): initial variances for student-specific offsets,
    \item \( \sigma^2_{\Delta \mu} \): innovation variance for the cohort trend,
    \item \( \boldsymbol{\sigma}^2_{\Delta \beta} := \{\sigma^2_{\Delta \beta_i}\} \): innovation variances for individual trajectories.
\end{itemize}

We place improper scale-invariant priors \( p(\sigma^2) \propto 1/\sigma^2 \) on the innovation variance terms, following the Jeffreys prior \citep{jeffreys}. This prior is widely used in hierarchical Bayesian models for its invariance under scale transformations and its flexibility in allowing the smoothness of latent trajectories to be learned from the data. It also enables efficient Gibbs sampling via conjugate inverse-gamma updates.  While improper and non-regularizing, this prior performs well when sufficient longitudinal data are available per individual \cite{gelman2006prior}, as is typically the case in our setting. By contrast, we place half-Cauchy priors with scale 1, $C^{+}(0,1)$, on the initial variance parameters $\sigma^2_{\mu}$ and $\sigma^2_{\beta}$, to provide regularization and support stable estimation at the first time point.

The full posterior is:
\[
p(\boldsymbol{\mu}, \boldsymbol{\beta}, \sigma^2_{\mu}, \sigma^2_{\Delta\mu}, \boldsymbol{\sigma}^2_{\beta}, \boldsymbol{\sigma}^2_{\Delta \beta} \mid \boldsymbol{y}).
\]

\paragraph{P\'{o}lya-Gamma data augmentation.}  
To address the non-conjugacy of the Bernoulli-logistic likelihood, we adopt the P\'{o}lya-Gamma (PG) data augmentation framework of \citet{pg}. Each observation likelihood can be re-expressed as:
\begin{align*}
    &f(y_{i,t,j}|\mu_t,\beta_{i,t},d_{j}) \\
    &= \frac{\exp\{(\mu_t + \beta_{i,t}) -d_{j}\}^{y_{i,t,j}}}{1+\exp\{(\mu_t + \beta_{i,t}) -d_{j}\}}.\\
    &\propto \int_{0}^{\infty}\exp\bigg\{\kappa_{i,t,j}((\mu_t + \beta_{i,t}) - d_{j})\bigg\} \\ 
    & \quad \quad \quad  \exp\bigg\{-\frac{\omega((\mu_t + \beta_{i,t}) -d_{j})^2}{2}\bigg\}p(\omega)d\omega,\\
    &\propto \int_{0}^{\infty} \mathcal{N}(\kappa_{i,t,j}|\omega(\theta_{i,t} + \beta_{i,t}- d_{j}), \omega) p(\omega)d\omega,
\end{align*}
where $\kappa_{i,t,j} = y_{i,t,j} - \frac{1}{2}$ and $\omega \sim PG(1,0).$ The P\'{o}lya-Gamma distribution with parameters $b>0$ and $c \in \mathcal{R}$, is denoted as PG(b,c), is defined as
\[
X \stackrel{D}{=}\frac{1}{2\pi^2}\sum_{k=1}^{\infty}\frac{g_k}{(k-1/2)^2+c^2/4\pi^2},
\]
where the $g_k \sim \text{Gamma}(b,1)$ and $\stackrel{D}{=}$ denotes equality in distribution. 

We exploit the banded structure of the random walk priors to perform efficient Gibbs sampling using the sparse Cholesky algorithm of \citet{rue}. Each iteration scales linearly in the number of students \( N \) and time steps \( T \). This structure, combined with the conjugacy induced by PG augmentation, enables exact posterior inference even in high-dimensional settings. PG-based samplers are also geometrically ergodic \citep{geoer}, providing theoretical guarantees for convergence.

\subsection{Comparison with alternative methods}  
Fully Bayesian inference offers calibrated uncertainty estimates, which are particularly valuable in sparse data settings. However, exact inference in logistic IRT models is challenging due to the non-conjugacy of the likelihood and the high dimensionality introduced by dynamic latent structures.

General-purpose samplers such as the No-U-Turn Sampler (NUTS) \citep{nuts}, implemented in Stan \citep{stan}, are widely used for models with complex posteriors due to their automatic tuning and robust convergence properties \citep{hmcgood,hmcbad3}. Yet these methods are often computationally infeasible for high-dimensional, structured time-series models like dynamic IRT due to poor scaling and slow mixing \citep{hmcbad1,hmcbad2}.

To improve scalability, many existing dynamic IRT models adopt approximate inference: \citet{wang_bayesian_2013} approximate the likelihood using a mixture-of-normals; \citet{imai2016fast} and \citet{kim2023variational} use variational inference. While efficient, these methods may introduce bias and understate posterior uncertainty.

In contrast, P\'{o}lya-Gamma augmentation enables exact posterior inference by transforming the logistic likelihood into a conditionally Gaussian form. This allows conjugate updates for latent trajectories and variance components, making it well-suited to dynamic IRT models like D-BIRD. Although less flexible than black-box or amortized inference approaches, PG-based Gibbs sampling provides a tractable, theoretically grounded alternative that supports full Bayesian inference at scale.

\section{Simulation Study}\label{sec:simulation}

\subsection{Design}
We conduct a simulation study to assess the parameter recovery performance of D-BIRD in comparison with two baselines:

\begin{itemize}
  \item \textbf{Global-RW:} No cohort trend; all students share the same innovation variance (analogous to the model specification used in \citet{kim2023variational, martin2002dynamic}): \label{eq:base1}
  \[
  \theta_{i,t} = \beta_{i,t}, \quad \Delta \beta_{i,t} \sim \mathcal{N}(0, \sigma^2_{\Delta\beta}).
  \]
  
  \item \textbf{Hetero-RW:} No cohort trend; each student has their own innovation variance: \label{eq:base2}
  \[
  \theta_{i,t} = \beta_{i,t}, \quad \Delta \beta_{i,t} \sim \mathcal{N}(0, \sigma^2_{\Delta\beta_i}).
  \]
\end{itemize}

\noindent This design allows us to assess how each feature improves recovery of latent ability trajectories and model parameters under controlled conditions.

We simulate response data for \( N = 150 \) students over \( T = 100 \) sessions, with 10 items per session. Ability is generated according to the D-BIRD specification (Equation~\ref{eq:fullmodel}), which includes both a global cohort trend \( \mu_t \) and individualized deviations \( \beta_{i,t} \). The cohort trend is simulated as a smooth Gaussian random walk:
\[
\mu_1 \sim \mathcal{N}(0, 0.1), \quad \Delta \mu_t \sim \mathcal{N}(0, 0.05).
\]
This latent trend is shared across all students and governs the population-wide evolution of ability.

To introduce heterogeneity in latent trajectories, we generate student-specific deviations $\beta_{i,t}$ with varying levels of smoothness. Students are split into two groups: the first 75 have low-variance random walks (more stable learning), while the remaining 75 have higher-variance trajectories:

\begin{align*}
&\beta_{i,t} = \hat{\beta}_{i,1} - \frac{1}{150}\sum_{i=1}^{150} \hat{\beta}_{i,t} \\
&\hat{\beta}_{i,1}\stackrel{iid}{\sim} \mathcal{N}(0, \sigma^2_{\beta_i}),\\ 
&\sigma^2_{\beta_i} \sim \mathrm{Gamma}(5,10),\\
&\Delta \hat{\beta}_{i,t} \stackrel{\text{iid}}{\sim} \mathcal{N}(0, \sigma^2_{\Delta \beta_i}), \\
&\sigma^2_{\Delta \beta_i} \sim
\begin{cases}
\text{Gamma}(5, 500), & \text{if } i \leq 75 \ (\text{Group A})\\
\text{Gamma}(5, 10),  & \text{if } i > 75 \ (\text{Group B}).
\end{cases}
\end{align*}

\begin{figure}[h]
  \centering
  \includegraphics[width=\linewidth]{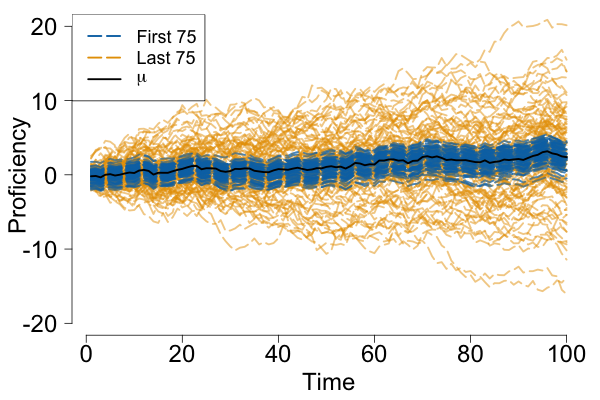}
  \caption{Simulated latent abilities $\theta_{i,t}$ for 150 students over 100 sessions. Orange lines represent the trajectories of the first 75 students; blue lines correspond to the remaining 75. The global trend shared across all students is shown in black.}
  \label{fig:simulated_theta}
\end{figure}

Figure~\ref{fig:simulated_theta} shows the simulated ability trajectories, where group differences in smoothness and the shared cohort pattern are visible. Item difficulties are drawn from $d_{i,j,t} \sim N(\theta_{i,t},0.5)$.

% Well-calibrated item difficulties are drawn from $d_{i,j,t} \sim N(\theta_{i,t},0.5)$. 

This design creates a data-generating process with two key properties: (1) a smooth global trajectory shared across all students, and (2) heterogeneous individual learning dynamics. D-BIRD is designed to exploit both sources of structure, while the baseline models can only recover one or the other. Each simulation is replicated 250 times, and recovery is evaluated using mean squared error (MSE), empirical coverage (EC), and mean credible interval width (MCIW).

\begin{table*}[t]
  \centering
  \begin{tabular}{lccc}
    \hline
    Model & MSE & EC & MCIW \\
    \hline
    \textbf{D-BIRD}         & \textbf{0.216 (0.008)} & \textbf{0.960 (0.004)} & \textbf{1.791 (0.03)} \\
    \textbf{Global-RW}  & 0.270 (0.011) & 0.944 (0.005) & 1.993 (0.054) \\
    \textbf{Hetero-RW}   & 0.260 (0.013) & 0.901 (0.038) & 1.801 (0.134) \\
    \hline
  \end{tabular}
  \caption{Posterior recovery metrics for student trajectories $\theta_{i,t}$, comparing our proposed model, D-BIRD, against two baselines, Global-RW and Hetero-RW. Metrics include mean squared error (MSE), empirical coverage (EC), and empirical credible interval width (ECIW), with standard deviations shown in parentheses.}
  \label{tab:recovery}
\end{table*}

\subsection{Results}
Table~\ref{tab:recovery} summarizes model performance across 250 replications. D-BIRD consistently achieves the lowest mean squared error (MSE), indicating superior accuracy in recovering latent ability trajectories. This reflects its ability to capture both the global trend and student-specific deviations—structure explicitly encoded in the data-generating process.

By contrast, the Global-RW model performs worst. Because it assumes a single shared innovation variance and lacks a cohort trend, it cannot accommodate the observed heterogeneity in trajectory smoothness across students. This mismatch leads to oversmoothing and inflated error, particularly for students with rapidly changing trajectories.

The Hetero-RW model improves on Global-RW by allowing individualized evolution variances. However, it treats each student's trajectory as independent, ignoring the shared global trend present in the data. As a result, it fails to borrow strength across students and exhibits higher estimation error than D-BIRD. In contrast, D-BIRD strikes a balance: it captures population-level structure via the cohort trend \( \mu_t \), while flexibly adapting to individual variation through student-specific deviations \( \beta_{i,t} \). This enables more stable and accurate recovery, especially in the presence of sparse data.

D-BIRD also outperforms both baselines in terms of uncertainty quantification. It achieves near-nominal empirical coverage (\textasciitilde96\%) with the narrowest credible intervals, as shown by the lowest MCIW. Hetero-RW exhibits undercoverage despite wide intervals, suggesting unstable variance estimation. Global-RW maintains nominal coverage but at the cost of overly wide intervals, due to its inability to represent individual variation. Overall, D-BIRD provides not only more accurate point estimates, but also sharper and more reliable posterior uncertainty.

\section{Empirical Application}
\label{sec:application}

We apply D-BIRD to longitudinal assessment data from a widely used digital K--12 learning platform to illustrate its practical utility. The goal is to show how the model recovers interpretable learning trajectories at both the cohort and individual levels over time. We also compare D-BIRD to static IRT estimates of ability, highlighting the added insight gained from dynamic modeling of student ability.

\subsection{Data and Setup}

Students on the platform begin with a full-length assessment comprising approximately 25 items drawn from a pre-calibrated Rasch item pool. Based on these initial estimates of ability, students are assigned a personalized instructional sequence, with each module followed by a brief 5-item quiz. Full-length assessments are re-administered periodically, providing updated proficiency estimates from static IRT and allowing for instructional adaptation. All item difficulties are known and expressed in Rasch logits.

In our analysis, we focus on two cohorts—Kindergarten (Grade 0) and Grade 5—to capture developmental contrasts in growth patterns. For both cohorts, we restrict the sample to students who completed at least four full-length assessments and truncate time series at 40 weeks. The final analytic sample includes 101 Kindergarten students and 311 Grade 5 students. For Kindergarten, the median observation span was 37 weeks, with a median of 19 active weeks and 10 responses per active week. For Grade 5, the median span was 39 weeks, with 20 active weeks and 14 responses per active week.

\subsection{Methods}
\begin{figure}[ht]
  \centering
  \includegraphics[width=0.8\linewidth]{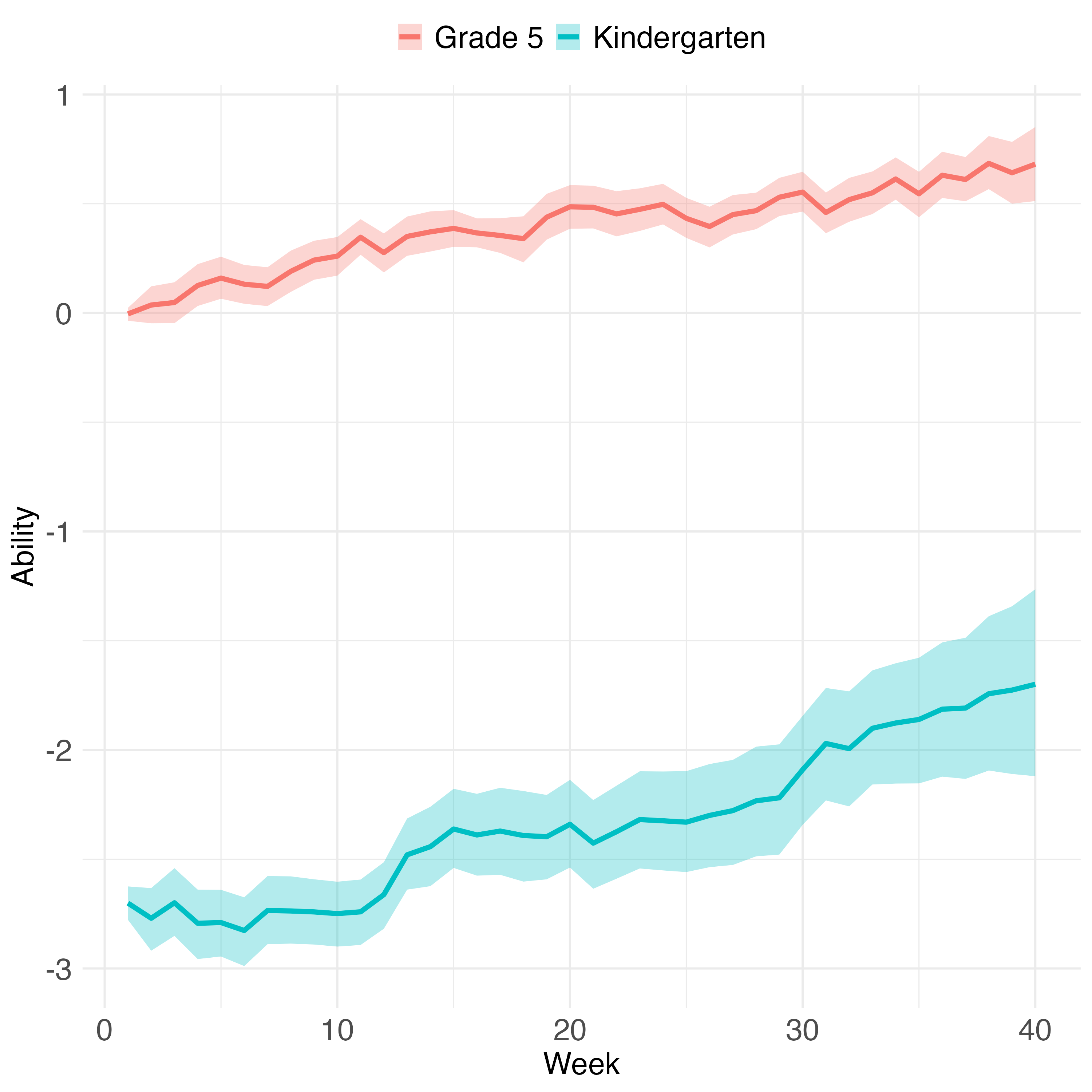}
  \caption{Estimated cohort-level ability trends \(\mu_t\) for Kindergarten and Grade 5. Bands show 95\% credible intervals.}
  \label{fig:mu-edmentum}
\end{figure}

\begin{figure*}[ht]
\centering
\begin{subfigure}[t]{0.48\textwidth}
\centering
\includegraphics[width=\linewidth]{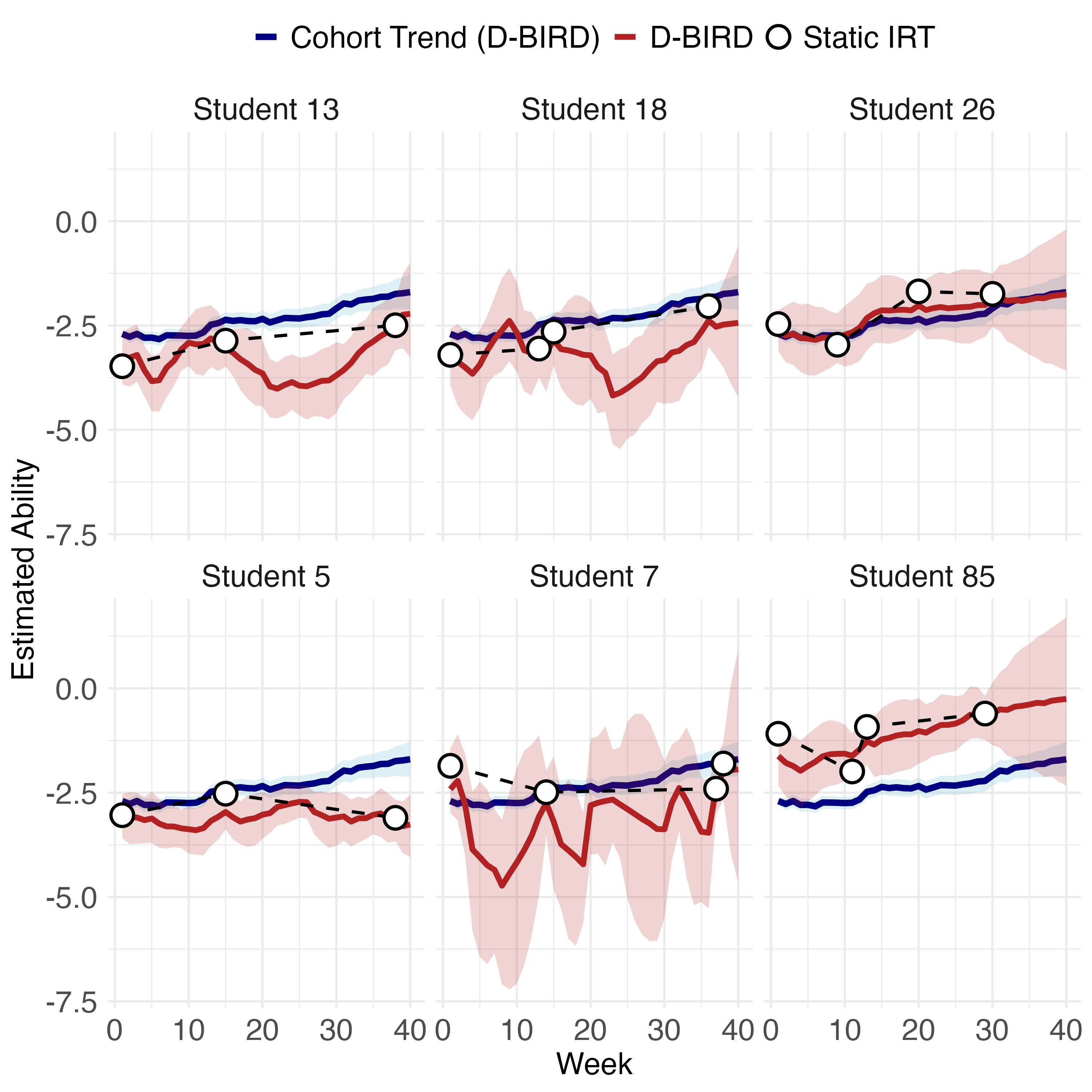}
\caption{Kindergarten}
\label{fig:grade0}
\end{subfigure}
\hfill
\begin{subfigure}[t]{0.48\textwidth}
\centering
\includegraphics[width=\linewidth]{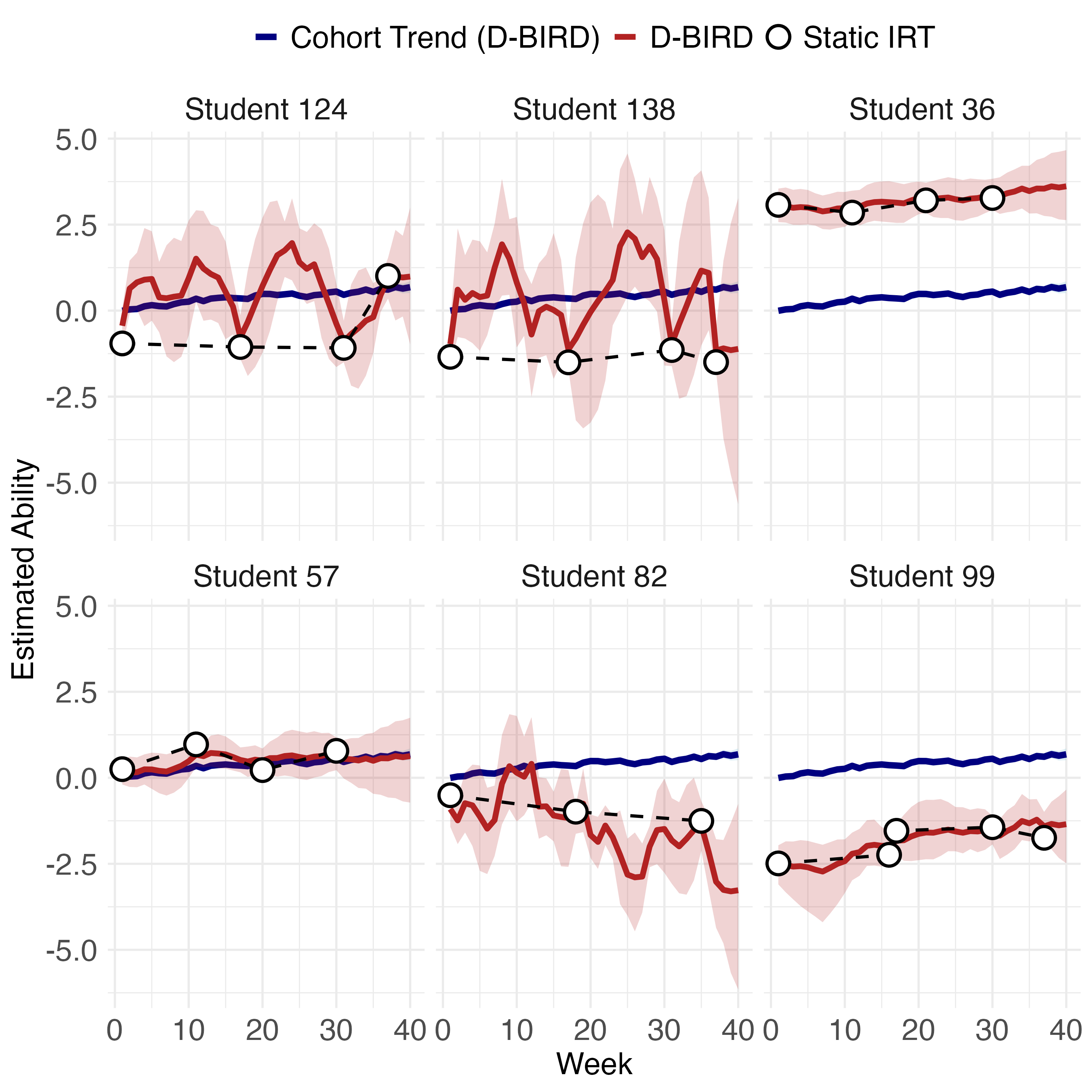}
\caption{Grade 5}
\label{fig:grade5}
\end{subfigure}
\caption{Estimated ability trajectories for selected students in Kindergarten (left) and Grade 5 (right). Red lines represent D-BIRD posterior means of ability with 95\% credible intervals; blue lines show estimated cohort trend with 95\% credible intervals. White circles indicate static IRT estimates from full-length assessments, connected by dashed lines for visual continuity (not model-derived).}
\label{fig:theta-edmentum}
\end{figure*}

To establish a static IRT baseline, we estimate each student's ability at the time of each full-length assessment using a Rasch model with the pre-calibrated item difficulties. Specifically, we compute the maximum a posteriori (MAP) estimate of ability under a logistic item response function and a Gaussian prior \( \theta \sim \mathcal{N}(0, 5^2) \). These estimates serve as snapshot summaries of student proficiency at irregular time points and are used for visual comparison with dynamic trajectories estimated by D-BIRD.

We then fit D-BIRD separately for each cohort, using the Bayesian inference procedure described in Section~\ref{sec:inference}. The model is estimated using 10{,}000 burn-in iterations followed by 10{,}000 posterior samples. We use the pre-calibrated item difficulties provided by the platform. D-BIRD yields posterior distributions for both the cohort-level trend \( \mu_t \) and the individual-specific deviations \( \beta_{i,t} \) at weekly resolution.

\subsection{Results}
\subsubsection{Cohort-Level Trends}
Figure~\ref{fig:mu-edmentum} shows the estimated cohort-level trends and their 95\% credible intervals over the 40-week period for Kindergarten and Grade 5 cohorts. As expected, Kindergarten students exhibit lower baseline ability ($\hat\mu_1^{\text{G0}} = -2.34$; 95\% CI: [-2.55, -2.12]) than Grade 5 students ($\hat\mu_1^{\text{G5}} = 0.39$; 95\% CI: [0.30, 0.49]). Kindergarten students exhibited steady growth (mean slope = 0.026 logits/week), while Grade 5 trends were flatter (mean = 0.018 logits/week), suggesting slower average gains. 

\subsubsection{Individual Ability Trajectories}

Figures~\ref{fig:grade0} and~\ref{fig:grade5} present D-BIRD ability trajectories for selected students in Kindergarten and Grade 5, respectively.

\paragraph{Kindergarten cohort.}
Students 26 and 85 both show upward trends in their static IRT scores, but D-BIRD reveals important distinctions. While Student 26 tracks closely with the cohort trend, Student 85 consistently outperforms it—something obscured without the group-level benchmark. In contrast, Student 5 appears to decline over time, falling further below the cohort average.

Static scores for Students 13, 18, and 26 appear similar at first glance, but D-BIRD uncovers meaningful differences in learning dynamics and uncertainty. Student 13 and 18 both show a mid-year dip, suggesting potential struggle despite an upward endpoint. Student 18’s wide posterior band reflects high uncertainty due to sparse data. Student 26 maintains steady growth in line with the cohort, highlighting the value of interpreting performance in temporal and contextual terms.

\paragraph{Grade 5 cohort.}
Students 36, 57, and 99 follow visually similar static score trajectories, yet D-BIRD differentiates them sharply when viewed against the cohort trend. Student 36 consistently outperforms the cohort while showing stable progress; Student 57 remains aligned with the cohort; and Student 99 lags well behind. These distinctions demonstrate how D-BIRD contextualizes student ability trajectories to the cohort trend.

Student 124 illustrates a different case. Their static scores remain low until a notable jump on the last full-length test. However, D-BIRD estimates their ability to have already increased in the weeks prior, indicating that quiz-level responses captured learning gains before they appeared in test scores. This fluctuating trajectory contrasts with the smoother paths of Students 36, 57, and 99, highlighting D-BIRD's sensitivity to between-test dynamics.

Finally, Students 82 and 138 both underperform on full-length tests, but their trajectories diverge. D-BIRD estimates a relatively stable, slightly declining path for Student 82, with a brief upward bump around week 10. Student 138, in contrast, shows more variability and potential mid-year recovery. These differences underscore D-BIRD’s ability to distinguish between superficially similar learners by leveraging the full sequence of assessment interactions.

\section{Discussion}
This paper introduces D-BIRD, a Bayesian dynamic IRT model that decomposes student ability into a shared cohort trend and an individual-specific trajectory. This structure is designed to support an important goal of educational measurement in personalized learning environments: tracking individual growth over time while situating it within broader group-level patterns. By explicitly modeling both individual and cohort dynamics, D-BIRD enables interpretable inferences even under sparse, irregular assessment conditions—a common feature of modern digital learning systems.

D-BIRD combines two key ideas: structured borrowing across students and flexible modeling of individual change. The cohort trajectory provides a stable, data-driven reference against which individual deviations can be interpreted. Student-specific innovation variances allow each learner’s ability to evolve with a level of smoothness appropriate to their observed responses. Exact Bayesian inference via P\'{o}lya-Gamma augmentation ensures well-calibrated posterior estimates, avoiding common approximations such as variational inference.

Several modeling choices limit the generalizability of D-BIRD and point to directions for future work. First, we assume item difficulties are known, consistent with operational settings that use pre-calibrated item pools. Future work could relax this assumption to jointly estimate item and ability parameters, exploring identifiability under sparsity. Second, D-BIRD is formulated in discrete time, where each time index may correspond to a learning opportunity \citep{koedinger2023astonishing}, a day of instruction \citep{wang_bayesian_2013}, or—as in our empirical application—a week. Extensions to continuous time, such as placing Gaussian process priors over latent ability \citep{8970762}, could support finer-grained modeling of learning dynamics, particularly in irregular data streams. Third, D-BIRD currently models dichotomous responses using the Rasch model. A natural extension is to adapt the framework for polytomous item models \citep{ostini2006polytomous}, enabling broader applicability to complex assessment formats. 

More broadly, D-BIRD contributes to a growing body of work at the intersection of psychometrics and AI-driven learning systems. As adaptive platforms increasingly rely on real-time data to personalize instruction, there is a pressing need for interpretable models that capture both individual learning progress and broader cohort-level trends. D-BIRD helps meet this need by offering a principled approach to longitudinal ability estimation—balancing flexibility with structure, and individual adaptation with population-level insight. In doing so, it advances longstanding goals in educational measurement while aligning with the practical demands of emerging digital learning environments.

\bibliography{main}

% \appendix
% \input{sections/appendix}

\end{document}